# A new class of exact solutions of the Klein-Gordon equation of a charged particle interacting with an electromagnetic plane wave in a medium

**Sándor Varró**
Wigner Research Centre for Physics, Hungarian Academy of Sciences,
Institute for Solid State Physics and Optics, Budapest, Hungary
E-mail: varro.sandor@wigner.mta.hu

**Abstract.** Exact solutions are presented of the Klein-Gordon equation of a charged particle moving in a classical monochromatic electromagnetic plane wave in a medium of index of refraction $n_m < 1$. The solutions are expressed in terms of Ince polynomials, which form a doubly infinite set labeled by two integer quantum numbers. These integer numbers represent quantized spectra of the momentum components of the charged particle along the polarization vector and along the propagation direction of the applied electromagnetic plane wave field. Since this field may represent a laser radiation of arbitrary high intensity propagating in an underdense plasma, the solutions obtained may have relevance, for instance, in describing possible quantum features of laser acceleration of electrons.

## 1. Introduction

In the theory of fundamental processes taking place in strong laser fields (Fedorov 1997) it is important to find exact, analytic solutions of the relativistic wave equations of charged particles, in order to consider their interaction beyond the perturbative regime. The best example for such solutions are the Volkov states of an electron or other charged particle, being the exact solutions of the Dirac equation (Volkov 1935), Klein-Gordon (Gordon 1927) or Schrödinger equation (Keldish 1964, Bunkin and Fedorov 1965), which have long been an important theoretical tool. In the relativistic regime the closed analytic form of this set of states rely on the vacuum dispersion relation of the classical electromagnetic (EM) plane wave representing the laser field. Soon after the advent of the laser, these solutions have been applied to describe various strong-field phenomena (see Brown and Kibble 1964, Eberly 1969, Neville and Rohrlich 1971, Gitman *et al* 1975, Ritus and Nikishov 1979, Bergou and Varró 1980), and their study has recently been undergoing a sort of 'renaissance' (see e.g. Panek *et al* 2002, Zakowicz 2005, Salamin *et al* 2006, Musakhayan 2008, Ehlotzky *et al* 2009, Boca and Florescu 2009, 2010, Boca 2011, Harvey *et al* 2012). The solutions of the Dirac and Klein-Gordon equation can also be expressed in a closed form for a quantized electromagnetic plane wave, as has first been shown by Bersons (1967). Besides the generalizations of these quantized states (Fedorov and Kazakov 1973, Bersons and Valmanis 1973), they have also been used for treating the nonlinear Compton scattering beyond the semiclassical approximation (Bergou and Varró 1981*c*). The generalization of the Volkov states with a quantized radiation have also been used to describe multiphoton stimulated Bremsstrahlung in the nonrelativistic regime (Bergou and Varró 1981*b*, Burenkov and Tikhonova 2010). The appearance of various versions of entanglement in strong-field laser-matter interactions has also been investigated by Fedorov *et al* (2006) for brakeup processes, and by Varró (2008, 2010*a*-*b*) for the photon-electron system itself. In the present paper we shall stay in the frame of the semiclassical description of the radiation field, and will not discuss many-particle correlations and entanglement in the presence of strong laser fields.

If the EM field propagates in a medium (with a real index of refraction $n_m \neq 1$), or the strong laser field is modelled by a time-periodic electric field, then the by now studied solutions of the relativistic wave equations can be expressed by the solutions of the corresponding Mathieu or Hill equations (see Narozhny and Nikishov 1974, Becker 1977, Cronström and Noga 1977). Such an analysis have for example been applied in the mathematical study of pair creation in strong fields (Nikishov and Ritus 1967, Nikishov1970). This phenomenon has recently received a renoved interest (see e.g. Narozhny *et al* 2004, Popov 2004, Dunne 2004, 2009), owing to the considerable technological development of extremely high power and ultrashort laser systems (see e.g. Krausz and Ivanov 2009, Mourou *et al* 2006 and 2013). Except for some special cases, the solutions cannot be expressed in a closed finite form; they are related to higher trancendental functions. We note that if one describes a Schrödinger particle, interacting with an EM plane wave in vacuum, beyond the dipole approximation (by taking exactly into account the spatial variation of the field), then one also encounters Mathieu or Hill equation, with their associated stability charts and band structures (Bergou and Varró 1981*a*). However, the occurrence of such band structures for the momenta has been shown to be a result of the inconsistent use of the nonrelativistic description for high-intensity fields, as has been pointed out by Drühl and McIver (1983).

In the present paper we show that there are exact closed form solutions of the Klein-Gordon equation of a charged particle moving in a monochromatic classical plane EM field in a medium of index of refraction $n_m < 1$. Since such radiation fields can be transformed to a homogeneous oscillating electric field (Narozhny and Nikishov 1974, Becker 1977), by going over to a suitably chosen Lorentz frame, our considerations are relevant for the discussion of this kind of interaction, too. The solutions to be derived are expressed in terms of finite trigonometric polynomials; the Ince polynomials whose arguments are integer or half-integer multiples of the EM wave's phase, and they form a doubly infinite discrete set of solutions labeled by two integer quantum numbers. This latter property is a completely new feature in comparison with both the Gordon-Volkov states (in vacuum) parametrized by continouos four momenta, and the Mathieu type solutions characterized by 'stability charts' resulting in a band structure of the allowed regions of the disposable parameters. We note that we have recently solved the analogous problem for a Dirac particle, too (Varró 2013).

In section 2, during the course of the derivation of Ince's equation from the Klein-Gordon equation, we also make a brief comparison with the method leading to the usual Volkov states or to the Mathieu-type wave functions. Section 3 is devoted to the determination of the real polynomial solutions, which are associated to the eigenvalue problem of special finite tri-diagonal matrices. The solutions coincide with the Ince polynomials, labeled by two integer quantum numbers, which represent a quantized spectrum of the electron's momentum components along the (linear) polarization direction and along the propagation direction of the laser field. In section 4 we shall summarize the main mathematical properties of the solutions obtained, and few physical implications associated to them will be mentioned. In general, more emphasize shall be placed on the existence and analytic form of these solutions, than on physical and numerical results. However, in section 4 some numerical examples will also be presented, just for illustration purposes. In section 5 a brief summary with conclusions closes our paper.

## 2. On the Volkov–type solutions and on the Mathieu–type solutions of the Klein-Gordon equation

In an external electromagnetic field characterized by the four-vector potential $A(x)$ the Klein-Gordon equation of a particle of charge $e$ and of mass $m$ has the form[1]

$$(\Pi^2-\kappa^2)\Phi=0\,,\qquad \Pi\equiv i\partial-\varepsilon A,\qquad \varepsilon=e/\hbar c\,,\qquad \kappa=mc/\hbar\,, \tag{1}$$

where $c$ is the velocity of light in vacuum, and $\hbar$ is Planck's constant divided by $2\pi$. In (1) we have introduced the operator of the kinetic four-momentum $\Pi$ of the charged particle. In a medium of index of refraction $n_m$, a general transverse electromagnetic plane wave of wave vector $k$ can be represented by a vector potential

$$A(x)=e_1A_1(\xi)+e_2A_2(\xi)\,,\quad \xi=k\cdot x\,,\quad k\cdot A=0\,,\quad k=\{k^\mu\}=k^0(1,n_m\boldsymbol{e}_3)\,,\quad \{e_{1,2}^\mu\}=(0,\boldsymbol{e}_{1,2})\,, \tag{2}$$

where $\{\boldsymbol{e}_1,\boldsymbol{e}_2,\boldsymbol{e}_3\}$ form a right system of mutually orthogonal unit vectors. In principle, the scalar functions $A_{1,2}(\xi)$ in (2) may have arbitrary form (satisfying, of course, certain regularity conditions). For a purely monochromatic field $A_{1,2}(\xi)$ are simple sine and cosine functions and $k^0=\omega_0/c$, where $\omega_0=2\pi\nu_0$ is the circular frequency. In case of a general plane EM field, the solutions can be expressed in a simple closed form (Gordon 1927). By taking a modified de Broglie plane wave, associated to a four-momentum parameter $p_\mu$, we have

$$\Phi = \Phi_p(\xi)\exp(-ip\cdot x),$$

$$-k^2\frac{d^2\Phi_p}{d\xi^2} + 2ik\cdot p\frac{d\Phi_p}{d\xi} + (p^2 - \kappa^2 - 2\varepsilon p\cdot A + \varepsilon^2 A^2)\Phi_p = 0, \qquad \xi = k\cdot x. \tag{3}$$

[1] The Minkowski metric tensor $g_{\mu\nu}=g^{\mu\nu}$ has the components $g_{00}=-g_{ii}=1$ $(i=1,\ 2,\ 3)$ and $g_{\mu\nu}=0$ if $\mu\neq\nu$ $(\mu,\nu=0,\ 1,\ 2,\ 3)$. The scalar product of two four-vectors $a$ and $b$ is $a\cdot b=g_{\mu\nu}a^\mu b^\nu$, i.e. $a\cdot b=a_\nu b^\nu=a^0b^0-\boldsymbol{a}\cdot\boldsymbol{b}$, where $\boldsymbol{a}\cdot\boldsymbol{b}$ is the usual scalar product of three-vectors $\boldsymbol{a}$ and $\boldsymbol{b}$. Space-time coordinates are denoted by $x^\mu$, where $x=\{x^\mu\}=(ct,\boldsymbol{r})$. The four-gradient is $\partial=\{\partial^\mu\}=(\partial/\partial ct,-\partial/\partial\boldsymbol{r})$, and it covariant components are $\partial_\mu=\partial/\partial x^\mu$.

In vacuum ($n_m = 1$, $k = k_0(1, \boldsymbol{e}_3)$, $k^2 = 0$) the factor of the second derivative in (3) is zero, and the remaining first order equation can be directly integrated, yielding the Gordon-Volkov states,

$$\Phi_p = (2\pi)^{-3/2}(2p_0)^{-1/2}\exp(-iS_p), \qquad S_p = p\cdot x + \frac{1}{2k\cdot p}\int[-2\varepsilon p\cdot A + \varepsilon^2 A^2]d\xi, \qquad k^2 = 0, \tag{4}$$

where $S_p$ is the classical action divided by $\hbar$. Equation (4) displays the generic structure of the Gordon-Volkov type states, the exact solutions of the Klein–Gordon equation of a charged particle interacting with a plane electromagnetic wave in vacuum. Concerning the basic mathematical properties of these solutions, see the early papers by Brown and Kibble (1964), Eberly (1969), Bergou and Varró (1980), and the recent work by Boca (2011).

The analysis becomes more complicated if one considers the interaction with a plane wave propagating in a medium of index of refraction $n_m \neq 1$. This is due to the presence of the non-vanishing second derivative in (4), since now $k^2 = k_0^2(1 - n_m^2) \neq 0$. In this case we can eliminate the first derivative in (4), or a priori use the following modified Ansatz at the outset,

$$\Phi = \Phi_p^{(\pm)}(\xi)\exp\left\{\mp i\left[p - \frac{(k\cdot p)}{k^2}k\right]\cdot x\right\} \equiv \Phi_p^{(\pm)}(\xi)\exp[\mp i(-\hat{p}\hat{x} - p_1x_1 - p_2x_2)], \qquad k^2 > 0, \tag{5a}$$

$$\hat{p} \equiv \hat{k}\cdot p/(k^2)^{1/2}, \qquad \hat{x} \equiv \hat{k}\cdot x/(k^2)^{1/2}, \qquad k = \{k^\mu\} = k^0(1, n_m\boldsymbol{e}_3), \qquad \hat{k} = \{\hat{k}^\mu\} = -k^0(n_m, \boldsymbol{e}_3), \tag{5b}$$

$$\frac{d^2\Phi_p^{(\pm)}}{d\xi^2} + \frac{1}{-k^2}\left[-p_\xi^2/k^2 + p^2 - \kappa^2 \mp 2\varepsilon p\cdot A + \varepsilon^2 A^2\right]\Phi_p^{(\pm)} = 0, \qquad p_\xi \equiv (k\cdot p), \tag{5c}$$

where the ambient sign $\mp$ refers to the 'positive and negativ energy solutions'. The plane wave factor in the first equation in (5a) has also been written as $\exp[\mp i(-\hat{p}\cdot\hat{x} - p_1x_1 - p_2x_2)]$, where $p_1 = e_1\cdot p$ and $p_2 = e_2\cdot p$ are the transverse momentum components. We have introduced the 'complementary wave vector' $\hat{k} = -k^0(n_m, \boldsymbol{e}_3)$ ($\hat{k}^2 = -k^2$ and $\hat{k}\cdot k = 0$), with the help of which one can derive

$$p = \frac{(k\cdot p)}{k^2}k - \frac{(\hat{k}\cdot p)}{k^2}\hat{k} - (p\cdot e_1)e_1 - (p\cdot e_2)e_2, \qquad \left[p - \frac{(k\cdot p)}{k^2}k\right]\cdot x = -\hat{p}\hat{x} - p_1x_1 - p_2x_2. \tag{5d}$$

If $n_m^2 < 1$, then $\hat{k}$ is space-like, and $\hat{p} \equiv \hat{k}\cdot p/(k^2)^{1/2}$ plays the role of a momentum component conjugated to the 'position variable' $\hat{x} \equiv \hat{k}\cdot x/(k^2)^{1/2}$ (see e.g. Becker 1977). Thus, the Ansatz in (5a) means a separation of variables, where $\hat{k}\cdot p$, $p_1$ and $p_2$ are 'three-momentum type parameters', and $p_\xi = (k\cdot p)$ may be said to be an 'energy type parameter'. This latter parameter $p_\xi$ does not explicitely show up in the solution, because $(p_\xi / k^2)k_\mu$ is subtracted from $p_\mu$ (see (5d)).

In case of a right-circularly polarized monochromatic plane EM wave, for instance, one has $A_{cir}(x) = A_0(e_x\cos\xi + e_z\sin\xi)$, $A_{cir}^2 = -A_0^2 =$constant, and one immediately obtains a *Mathieu equation* for the modulation function $\Phi_p^{(\pm)}$, which we write in the standard form (Meixner and Schäfke 1954)

$$\frac{d^2w}{dz^2} + (\lambda - 2h^2\cos 2z)w = 0\,, \quad w \equiv \Phi_p^{(\pm)}\,, \quad \lambda \equiv (2k\cdot p)^2/(k^2)^2\,, \quad h^2 \equiv \mp 4p_\perp \varepsilon A_0 / k^2\,, \tag{6a}$$

$$z = (\xi - \alpha)/2\,, \quad \xi = \omega_0(t - n_m y/c)\ \sin\alpha = p_z / p_\perp\,, \quad p_\perp = \sqrt{p_x^2 + p_z^2}\,, \quad p^2 = \kappa^2 + \varepsilon^2 A_0^2\,. \tag{6b}$$

(see e.g. Becker 1977, Cronström and Noga 1977). We note that the two terms $\kappa^2$ and $\varepsilon^2 A_0^2$ in the mass-shell relation are usually combined to $\kappa_*^2 \equiv \kappa^2 + \varepsilon^2 A_0^2$, which is the intensity-dependent mass shift $\Delta m = m_* - m = m\sqrt{1+\mu_0^2} - m$, where $\mu_0 = eF_0 / mc\omega_0$ is the well-known dimensionless intensity parameter (see e.g. Brown and Kibble 1964, and the recent work by Harvey *et al* 2012).

Henceforth, in equation (1) and in (5c) we specialize $A = A(\xi)$ to represent a monochromatic linearly polarized plane wave,

$$A(x) = e_x A_0\cos\xi\,, \quad \xi = k\cdot x\,, \quad \{k^\mu\} = \frac{\omega_0}{c}(1,\ 0,\ n_m,\ 0)\,, \quad \{e_x^\mu\} = (0,\ 1,\ 0,\ 0)\,, \quad A_0 = F_0 / k_0\,, \tag{7}$$

where $\omega_0 = 2\pi\nu_0$ is the circular frequency and $F_0$ denotes the amplitude of the electric field strength. This is an $x-$polarized wave which propagates in the positive $y-$direction in the medium, i.e. the explicit form of the argument of the cosine is $\xi = k\cdot x = \omega_0(t - n_m y/c)$. The differential equation (5c) for modulation function $\Phi_p^{(\pm)}$ can be brought to the form,

$$\frac{d^2\Phi_p^{(\pm)}}{dz^2} + (\theta_0 + 2\theta_1\cos 2z + 2\theta_2\cos 4z)\Phi_p^{(\pm)} = 0\,, \quad z = \xi/2\,, \tag{8a}$$

$$\theta_0 \equiv \frac{4}{-k^2}[-(k\cdot p)^2/k^2 + p^2 - \kappa^2 - \varepsilon^2 A_0^2/2]\,, \quad 2\theta_1 \equiv \frac{4}{-k^2}[\pm 2p_x\varepsilon A_0]\,, \quad 2\theta_2 \equiv \frac{4}{-k^2}[-\varepsilon^2 A_0^2/2]\,. \tag{8b}$$

In (8a) we have defined the independent variable $z = \xi/2$ (as is usual in the theory of differential equations with periodic coefficients), and followed the standard notations for the parameters $\theta_{0,1,2}$. According to (8a), in case of a linearly polarized monochromatic wave (7), the modulation function $\Phi_p^{(\pm)}(\xi)$ satisfies the so-called *Whittaker–Hill equation* (or Hill's three-term equation), because now $A^2 = -A_0^2(1+\cos 2\xi)/2 \neq$ constant. If one attempts to solve this equation in terms of a trigonometric series, that procedure leads to five-term recurrence relations between the coefficients, in contrast to the three-term expressions encountered with the Mathieu equation. Thus, the standard procedure (the technique of continued fractions) cannot be taken over from the theory of Mathieu equation. In looking for finite-term solutions, we overcome this difficulty by using the transformation, originally due to Ince (see Arscott 1964, and references therein). This is the basic new element in our present approach. In the spirit of this method, in order to solve (8a), we proceed by introducing the following Ansatz for $\Phi_{ps}^{(\pm)}$

$$\Phi_p^{(\pm)} = w\exp(-\theta_2^{1/2}\cos 2z), \quad \frac{d^2w}{dz^2} + 4\theta_2^{1/2}\sin 2z\frac{dw}{dz} + [\theta_0 + 2\theta_2 + (2\theta_1 + 4\theta_2^{1/2})\cos 2z]w = 0\,. \tag{9}$$

By using a more condensed notation, from (9) we receive the standard form of *Ince's equation*,

$$\frac{d^2w}{dz^2} + a\sin 2z\frac{dw}{dz} + (\eta - qa\cos 2z)w = 0, \quad z = \xi/2, \quad a \equiv 4\theta_2^{1/2} = 4|\varepsilon|A_0/k_p, \tag{10a}$$

$$2\theta_1 \equiv -(q+1)a \to 2p_x = (q+1)k_p, \quad k_p \equiv \sqrt{k^2} = k_0\sqrt{1-n_m^2} \equiv \omega_p/c, \tag{10b}$$

$$\eta \equiv \theta_0 + 2\theta_2 = 4[p_\xi^2/k_p^2 + \varepsilon^2A_0^2 + \kappa^2 - p^2]/k_p^2, \quad \eta = 4[\hat{p}^2 + \mathbf{p}_\perp^2 + \kappa^2 + \varepsilon^2A_0^2]/k_p^2. \tag{10c}$$

The second equation of (10c) from the first one has been derived by using the relations

$$p_0^2 - p_y^2 = p_\xi^2/k_p^2 - \hat{p}^2, \quad p^2 = p_0^2 - p_y^2 - \mathbf{p}_\perp^2 = p_\xi^2/k_p^2 - \hat{p}^2 - \mathbf{p}_\perp^2, \tag{10d}$$

which is valid for an arbitrary four vector $p_\mu$, where $\hat{p}$ and $p_\xi$ have been defined in (5b) and (5c), respectively. In obtaining (10a-b-c) from (8a-b) we have taken a negative test charge (electron; $\varepsilon < 0$), and considered 'positive solutions'. Note that equation (10a) is unchanged under the simultaneous substitutions $z \to z + \pi/2$ and $a \to -a$, thus we need not consider the case $\varepsilon > 0$ and $\varepsilon < 0$ separately. In (10b) we have introduced the new parameters $q$ and $k_p \equiv k_0\sqrt{1-n_m^2} \equiv \omega_p/c$, where the subscript '$p$' in the latter symbol refers to the word 'plasma', though at the present stage we do not need to specify the nature of the medium (in section 4 we shall deal with this interpretation).

In the rest of the present paper we shall study the general equation (10a), however, we shall not carry out a complete analysis of this equation, which has a quite unexplored infinite set of transcendental solutions. Our analysis is restricted to the special class of solutions, the (finite) Ince polynomials, which form a subset of all solutions. The exceptional feature of these solutions is that they form a doubly infinite countable set, corresponding to discrete values of the transverse and longitudinal momentum parameters.

## 3. Periodic, finite solutions of the Klein-Gordon equation

In the previous section we have derived Ince's equation (10a) for the modulation function of a de Broglie plane wave solution of the original Klein-Gordon equation (1). In order to obtain periodic finite solutions we follow the procedure explained in chapter 7 of Arscott (1964). At this point we would like to note that, we have already used an analogous method for obtaining a new class of exact solutions of the corresponding Dirac equation, where, in contast to (10a), we have encountered a complex equation, whose solutions turned out to be finite complex Fourier sums (Varró 2013). Though the method itself used there is analogous to that we are now using, the complex wave functions obtained have quite different properties to that of the real Ince polynomials, which we are deriving in the present section.

Let us first expand the solution of Ince's equation (10a) as a cosine series,

$$w = \sum_{r=0}^{\infty} A_r \cos 2rz \,. \tag{11}$$

After substitution (10a), we have the recurrence relations for the unknown coefficients,

$$-\eta A_0 + (\tfrac{1}{2}q + 1)aA_1 = 0\,, \tag{11a}$$

$$qaA_0 + (4-\eta)A_1 + (\tfrac{1}{2}q + 2)aA_2 = 0\,, \tag{11b}$$

$$(\tfrac{1}{2}q - r)aA_r + [4(r+1)^2 - \eta]A_{r+1} + (\tfrac{1}{2}q + r + 2)aA_{r+2} = 0\,. \tag{11c}$$

If $q$ is a positive even integer, $q = 2n$, and if moreover $\eta$ is chosen that $A_{n+1} = 0$, then, according to (11c) with $r = n$ we have $A_{n+2} = 0$ also, thus, successively $A_{n+3} = A_{n+4} = \ldots = 0$. This means that the series terminates with the term $A_n \cos 2nz$, and (11a-b-c) becomes

$$-\eta A_0 + (n+1)aA_1 = 0\,, \tag{12a}$$

$$2naA_0 + (4-\eta)A_1 + (n+2)aA_2 = 0\,, \tag{12b}$$

$$(n-r)aA_r + [4(r+1)^2 - \eta]A_{r+1} + (n+r+2)aA_{r+2} = 0\,. \tag{12c}$$

The solution of this system is corresponds to a standard algebraic eigenvalue problem of the following tri-diagonal $(n+1)\times(n+1)$ matrix.

$$\begin{bmatrix} 0 & (n+1)a & & & \\ 2na & 4 & (n+2)a & & \\ & (n-1)a & 4\cdot 2^2 & \ldots & \\ & & \ldots & \ldots & 2na \\ & & & (n-(n-1))a & 4n^2 \end{bmatrix} \cdot \begin{bmatrix} A_0 \\ A_1 \\ \vdots \\ \vdots \\ A_n \end{bmatrix} = \eta \cdot \begin{bmatrix} A_0 \\ A_1 \\ \vdots \\ \vdots \\ A_n \end{bmatrix}, \quad \boldsymbol{L}_n(a)\boldsymbol{A} = \eta \boldsymbol{A}\,. \tag{12d}$$

Following Arscott's notation, we use the symbol for the tri-diagonal matrix,

$$\boldsymbol{L}_n \equiv \begin{bmatrix} \beta_0 & \gamma_0 & & & & \\ \alpha_1 & \beta_1 & \gamma_1 & & & \\ & \alpha_2 & \beta_2 & \gamma_2 & & \\ & & \ddots & \ddots & \ddots & \\ & & & \ddots & \ddots & \gamma_{n-1} \\ & & & & \alpha_n & \beta_n \end{bmatrix} = \left\langle \begin{matrix} & \gamma_0 & .... & .... & \gamma_{n-1} \\ \beta_0 & \beta_1 & .... & .... & \beta_n \\ \alpha_1 & \alpha_2 & .... & \alpha_n & \end{matrix} \right\rangle . \tag{13}$$

With this notation the tri-diagonal matrix, $\boldsymbol{L}_n(a)$ in (12d) is represented by the following symbol

$$\left\langle \begin{matrix} & (n+1)a & (n+2)a & ......... & 2na \\ 0 & 4 & 16 & ......... & 4n^2 \\ 2na & (n-1)a & (n-2)a & .... \ a & \end{matrix} \right\rangle = \boldsymbol{L}_n(a), \qquad \lambda_n(\eta,a) = \det[\boldsymbol{L}_n(a) - \eta \boldsymbol{I}]. \tag{14}$$

In (14) we have also introduced the characteristic polynomial $\lambda_n(\eta,a)$ associated to the eigenvalue problem, where $\boldsymbol{I}$ is the $(n+1)\times(n+1)$ unit matrix. The eigenvalues are given by the zeros of the characteristic polynomial $\lambda_n(\eta,a) = \prod_{k=0}^{n}(\eta - \eta_n^{(k)})$, in our case they are all real, and they are all simple roots of the equation $\lambda_n(\eta,a) = 0$. This is a consequence of the following lemma quoted on page 21 by Arscott (1964): „LEMMA 1.(a) *Let the numbers* $\alpha_i$, $\beta_i$, $\gamma_i$ *be all real, each product* $\alpha_{i+1}\gamma_i$ *positive, and let* $\mathbf{L}_n$ *denote the the tri-diagonal matrix above* [(13)], *all other elements being zero. Then the latent roots of* $\mathbf{L}_n$ *are all real and different.*” (We have inserted „[(13)]” into the quotation, in order to refer our equation (13) above.) There are $(n+1)$ real linearly independent eigenvectors $\{A_r^{(k)}\}$, associated to the eigenvalues $\eta_n^{(k)}$ $(0 \le k \le n)$. According to the above considerations, if $q = 2n$, i.e. if $p_x = k_p(n + 1/2)$, then the potentially infinite cosine series in (11) reduces to the Ince polynomial

$$w = \varphi_n^k(\xi \mid a, \mathrm{c}) = \sum_{r=0}^{n} A_r^{(k)}(a \mid n+1)\cos 2rz, \qquad z = \xi/2, \qquad (n = 1,\ 2,\ ...), \qquad (0 \le k \le n). \tag{15}$$

These solution may be called ‘even solution’ of (10a). In the notation $\varphi_n^k(\xi \mid a, \mathrm{c})$ the letter "c" refers to the word „cosine”, and in $A_r^{(k)}(a \mid n+1)$ the second argument refers to that there are $n+1$ sets of the coefficients (eigenvectors), and the superscript labels the $k$ – th such set.

We note that Ince introduced the following notations for the ordered roots of $\lambda_n(\eta,a) = 0$ and for the associated polynomials (Arscott 1964)

$$\eta = \eta_n^{(m)} \equiv a_{2n}^{2m}(a) \quad (m = 0,1,...,n), \qquad a_{2n}^0 < a_{2n}^2 < ... < a_{2n}^{2n}, \qquad w = C_{2n}^{2m}(z,a) = \sum_{r=0}^{n} A_r \cos 2kz . \tag{16a}$$

These functions are normalized and they are orthogonal to each other,

$$\int_{-\pi}^{\pi}[C_{2n}^{2m}(z,a)]^2\,dz=\pi\,,\qquad \int_{-\pi}^{\pi}e^{-(a/2)\cos 2z}C_{2n}^{2l}(z,a)C_{2n}^{2m}(z,a)dz=0\,,\qquad l\neq m\,. \tag{16b-c}$$

The latter orthogonality relation can be derived in the usual way, by using (8a), (9) and (10a). The wave function in equation (15) are the Ince polynomial $C_{2n}^{2k}(z,a)$ displayed in (16a),

$$w=\varphi_n^k(\xi\,|\,a,\mathrm{c})=\sum_{r=0}^{n}A_r^{(k)}(a\,|\,n+1)\cos 2rz=C_{2n}^{2k}(z,a)\,. \tag{16d}$$

Still for $q$ an even positive number, $q=2n$, there are also odd solutions of Ince's equation (10a). Let us now expand the wave function as a series of sines,

$$w=\sum_{r=0}^{\infty}B_r\sin 2rz\ , \tag{17}$$

The recurrence relations for the unknowns have the same structure as Eq. (12a-b-c), except that the first is omitted, and we take $B_0=0$,

$$(4-\eta)B_1+(\tfrac{1}{2}q+2)aB_2=0\,, \tag{17a}$$

$$(\tfrac{1}{2}q-r)aB_r+[4(r+1)^2-\eta]B_{r+1}+(\tfrac{1}{2}q+r+2)aB_{r+2}=0\ . \tag{17b}$$

If $q$ is a positive even integer, $q=2n$, and if moreover $\eta$ is chosen that $B_{n+1}=0$, then, according to Eq. (17b) with $r=n$ we have $B_{n+2}=0$ also, thus, successively $B_{n+3}=B_{n+4}=...=0$, which means that the series terminates with the term $B_n\sin 2nz$,

$$(4-\eta)B_1+(n+2)aB_2=0\,, \tag{18a}$$

$$(n-r)aB_r+[4(r+1)^2-\eta]B_{r+1}+(n+r+2)aB_{r+2}=0\,. \tag{18b}$$

The solutions of this system are determined by solving the eigenvalue problem of the $n\times n$ tri-diagonal matrix

$$\begin{bmatrix} 4 & (n+2)a & & & \\ (n-1)a & 4\cdot 2^2 & (n+3)a & & \\ & (n-2)a & 4\cdot 3^2 & \cdots & \\ & & \cdots & \cdots & (n+n)a \\ & & & (n-(n-1))a & 4n^2 \end{bmatrix}\cdot\begin{bmatrix} B_1\\ B_2\\ \vdots\\ \vdots\\ B_n\end{bmatrix}=\eta\cdot\begin{bmatrix} B_1\\ B_2\\ \vdots\\ \vdots\\ B_n\end{bmatrix},\qquad \boldsymbol{M}_n(a)\boldsymbol{B}=\eta\boldsymbol{B}\,. \tag{18c}$$

By using the condensed notation, as in (14), we write

$$\left\langle\begin{matrix} & (n+2)a & ........ & 2na \\ 4 & 4\cdot 2^2 & ........ & 4n^2 \\ (n-1)a & (n-2)a & .... & a \end{matrix}\right\rangle=\boldsymbol{M}_n(a)\,,\qquad \mu_n(\eta,a)=\det[\boldsymbol{M}_n(a)-\eta\boldsymbol{I}]\,, \tag{19}$$

where $\boldsymbol{I}$ means here the $n\times n$ unit matrix. The eigenvalues are the roots of the characteristic polynomial, i.e. $\mu_n(\eta_n^{(k)},a)=0$. The matrix $\boldsymbol{M}_n(a)$ in (19) also satisfies the conditions of the above lemma, thus the eigenvalues are all real and simple. In this way, we have found that, if $q = 2n$, i.e. if $p_x = k_p(n + \frac{1}{2})$, then the potentially infinite sine series in (17) reduces to the following Ince polynomial

$$w = \varphi_n^k(\xi \mid a, \mathrm{s}) = \sum_{r=1}^{n} B_r^{(k)}(a \mid n) \sin 2rz , \quad z = \xi / 2 , \quad (n = 1,\ 2,\ ...), \quad (1 \le k \le n) . \tag{20}$$

These solution may be called 'odd solution' of (10a). In the notation $\varphi_n^k(\xi \mid a, \mathrm{s})$ the letter "s" refers to the word „sine", and in the coefficients $B_r^{(k)}(a \mid n)$ the second argument shows that there are $n$ sets of the coefficients (eigenvectors), where the superscript labels the $k-$th such set. The properties of the found 'even sine-type solutions' have been summarized by Arscott (1964) in complete analogy with (16a-b-c),

$$\eta = \eta_n^{(m)} \equiv b_{2n}^{2m}(a) \quad (m = 1,2,...,n), \quad b_{2n}^2 < b_{2n}^4 < ... < b_{2n}^{2n} , \quad w = S_{2n}^{2m}(z,a) = \sum_{r=1}^{n} B_r \sin 2rz , \tag{21a}$$

$$\int_{-\pi}^{\pi} [S_{2n}^{2m}(z,a)]^2 dz = \pi , \quad \int_{-\pi}^{\pi} e^{-(a/2)\cos 2z} S_{2n}^{2l}(z,a) S_{2n}^{2m}(z,a) dz = 0 , \quad l \ne m . \tag{21b-c}$$

The 'even sine-type solutions' in (20) thus coincide with the corresponding Ince polynomials in (21a),

$$w = \varphi_n^k(\xi \mid a, \mathrm{s}) = \sum_{r=0}^{n} B_r^{(k)}(a \mid n) \sin 2rz = S_{2n}^{2k}(z,a) . \tag{21d}$$

There are finite solutions with $q$ being an odd number. Let us try to find these in the form

$$w = \sum_{r=0}^{\infty} A'_r \cos(2r+1)z . \tag{22}$$

The coefficients satisfy the recurrence relations

$$[\tfrac{1}{2}(q+1)a - (\eta - 1)]A'_0 + \tfrac{1}{2}(q+3)aA'_1 = 0 , \tag{22a}$$

$$\tfrac{1}{2}[q - (2r+1)]aA'_r + [(2(r+1)+1)^2 - \eta]A'_{r+1} + \tfrac{1}{2}[q + (2(r+2)+1)]aA'_{r+2} = 0 . \tag{22b}$$

Let us assume that $q$ is a positive odd integer, $q = 2n+1$, and require that $\eta$ is such that at $r = n+1$ the coefficient $A'_{n+1} = 0$ vanishes. Then, by putting $r = n$ in (22b) we see that the first two terms on the left hand side are zero, consequently $A'_{n+2} = 0$ vanishes, too, and all the higher index coefficient vanish, thus we have again a finite term expression for the wave function. Accordingly, (22a-b) are rewritten as

$$[-\eta + 1 + (n+1)a]A'_0 + (n+2)aA'_1 = 0 , \tag{23a}$$

$$(n-r)aA'_r + [(2(r+1)+1)^2 - \eta]A'_{r+1} + (n+1+r+2)aA'_{r+2} = 0 . \tag{23b}$$

The characteristic numbers are solutions of the eigenvalue problem of the $(n+1)\times(n+1)$ tri-diagonal matrix

$$\begin{bmatrix} 1+(n+1)a & (n+2)a & & & \\ na & 3^2 & (n+3)a & & \\ & (n-1)a & 5^2 & \dots & \\ & & \dots & \dots & (2n+1)a \\ & & & a & (2n+1)^2 \end{bmatrix} \cdot \begin{bmatrix} A'_0 \\ A'_1 \\ \vdots \\ \vdots \\ A'_n \end{bmatrix} = \eta \cdot \begin{bmatrix} A'_0 \\ A'_1 \\ \vdots \\ \vdots \\ A'_n \end{bmatrix}, \quad \boldsymbol{K}_n(a)\boldsymbol{A}' = \eta \boldsymbol{A}' . \tag{23c}$$

By using the condensed notation introduced in (13), we have

$$\left\langle \begin{matrix} & (n+2)a & (n+3)a & \dots\dots\dots & (2n+1)a \\ 1+(n+1)a & 3^2 & 5^2 & \dots\dots\dots & (2n+1)^2 \\ na & (n-1)a & (n-2)a & \dots. \ a & \end{matrix} \right\rangle = \boldsymbol{K}_n(a) , \qquad \kappa_n(\eta, a) = \det[\boldsymbol{K}_n(a) - \eta \boldsymbol{I}] , \tag{24}$$

where $\boldsymbol{I}$ means the $(n+1)\times(n+1)$ unit matrix. According to the above lemma, the eigenvalues (the roots of the characteristic equation $\kappa_n(\eta,a)=0$) are all real and different. We have shown that if $q=2n+1$, i.e. if $p_x = k_p(n+1)$, then the potentially infinite cosine series in (22) reduces to the Ince polynomial

$$w = \chi_n^k(\xi \mid a, \mathrm{c}) = \sum_{r=0}^{n} A_r'^{(k)}(a \mid n+1)\cos(2r+1)z , \qquad z = \xi/2 , \qquad (n = 0,\ 1,\ \dots) , \qquad (0 \le k \le n) . \tag{25}$$

By using Ince's notation, the ordered roots of $\kappa_n(\eta,a)=0$ and the associated polynomials are

$$\eta = \eta_n^{(m)} \equiv a_{2n+1}^{2m+1}(a) \ \ (m = 0,1,\dots,n), \quad a_{2n+1}^{1} < a_{2n+1}^{2+1} < \dots < a_{2n+1}^{2n+1} , \quad w = C_{2n+1}^{2m+1}(z,a) = \sum_{r=0}^{n} A'_r \cos(2r+1)z . \tag{26a}$$

The normalization condition and the orthogonality relation of $C_{2n+1}^{2k+1}(z,a)$ are the same as in (16b-c),

$$\int_{-\pi}^{\pi} [C_{2n+1}^{2m+1}(z,a)]^2 dz = \pi , \qquad \int_{-\pi}^{\pi} e^{-(a/2)\cos 2z} C_{2n+1}^{2l+1}(z,a) C_{2n+1}^{2m+1}(z,a) dz = 0 , \qquad l \ne m . \tag{26b-c}$$

The wave function in equation (25) can be expressed with the Ince polynomial $C_{2n+1}^{2k+1}(z,a)$ in (26a),

$$w = \chi_n^k(\xi \mid a, \mathrm{c}) = \sum_{r=0}^{n} A_r'^{(k)}(a \mid n+1)\cos(2r+1)z = C_{2n+1}^{2k+1}(z,a) . \tag{26d}$$

Finally we study the 'odd sine–type solutions'

$$w = \sum_{r=0}^{\infty} B'_r \sin(2r+1)z . \tag{27}$$

$$[-\eta + 1 - \tfrac{1}{2}(q+1)a]B'_0 + \tfrac{1}{2}(q+3)aB'_1 = 0 , \tag{27a}$$

$$\tfrac{1}{2}[q-(2r+1)]aB'_r + [(2(r+1)+1)^2 - \eta]B'_{r+1} + \tfrac{1}{2}[q+(2(r+2)+1)]aB'_{r+2} = 0 . \tag{27b}$$

By using the same reasoning as above for an odd $q = 2n+1$, from (27a-b) we have

$$[-\eta + 1 - (n+1)a]B'_0 - (n+2)aB'_1 = 0 , \tag{28a}$$

$$(n-r)aB'_r + [(2(r+1)+1)^2 - \eta]B'_{r+1} + (n+1+r+2)aB'_{r+2} = 0 , \tag{28b}$$

$$\begin{bmatrix} 1-(n+1)a & (n+2)a & & & \\ na & 3^2 & (n+3)a & & \\ & (n-1)a & 5^2 & \dots & \\ & & \dots & \dots & (2n+1)a \\ & & & a & (2n+1)^2 \end{bmatrix} \cdot \begin{bmatrix} B'_0 \\ B'_1 \\ \vdots \\ \vdots \\ B'_n \end{bmatrix} = \eta \cdot \begin{bmatrix} B'_0 \\ B'_1 \\ \vdots \\ \vdots \\ B'_n \end{bmatrix}, \quad \widetilde{\boldsymbol{K}}_n(a)\boldsymbol{B}' = \eta \boldsymbol{B}'. \tag{28c}$$

With the condensed notation introduced in (13), we have

$$\left\langle \begin{matrix} & (n+2)a & (n+3)a & \dots\dots\dots & (2n+1)a \\ 1-(n+1)a & 3^2 & 5^2 & \dots\dots\dots & (2n+1)^2 \\ na & (n-1)a & (n-2)a & \dots. & a \end{matrix} \right\rangle = \widetilde{\boldsymbol{K}}_n(a), \quad \widetilde{\kappa}_n(\eta,a) = \det[\widetilde{\boldsymbol{K}}_n(a) - \eta \boldsymbol{I}], \tag{29}$$

where $\boldsymbol{I}$ is the $(n+1)\times(n+1)$ unit matrix. As has been pointed out by Arscott (1964), the characteristic equation $\widetilde{\kappa}_n(\eta,a)=0$ is equivalent to $\kappa_n(\eta,-a)=0$, where $\kappa_n(\eta,a)$ has been defined in (24). To summarize this part of the section, we have found that if $q=2n+1$, i.e. if $p_x = k_p(n+1)$, then the cosine series in (22) reduces to the Ince polynomial

$$w = \chi_n^k(\xi \mid a,\mathrm{s}) = \sum_{r=0}^{n} B_r'^{(k)}(a \mid n+1)\sin(2r+1)z, \quad z=\xi/2, \quad (n=0,\ 1,\ \dots), \quad (0 \le k \le n). \tag{30}$$

By using Ince's notation, the ordered roots of $\widetilde{\kappa}_n(\eta,a)=0$ and the associated polynomials are

$$\eta = \eta_n^{(m)} \equiv b_{2n+1}^{2m+1}(a) \ \ (m=0,1,\dots,n), \ \ b_{2n+1}^{1} < b_{2n+1}^{2+1} < \dots < b_{2n+1}^{2n+1}, \ w = S_{2n+1}^{2m+1}(z,a) = \sum_{r=0}^{n} B_r'^{(m)}\sin(2r+1)z \ . \tag{31a}$$

The normalization condition and the orthogonality relation of $S_{2n+1}^{2m+1}(z,a)$ are the same as in (26b-c),

$$\int_{-\pi}^{\pi} [S_{2n+1}^{2m+1}(z,a)]^2 dz = \pi, \qquad \int_{-\pi}^{\pi} e^{-(a/2)\cos 2z} S_{2n+1}^{2l+1}(z,a) S_{2n+1}^{2m+1}(z,a) dz = 0, \qquad l \ne m. \tag{31b-c}$$

The wave function in equation (30) can be expressed with the Ince polynomial $S_{2n+1}^{2k+1}(z,a)$ in (31a),

$$w = \chi_n^k(\xi \mid a,\mathrm{s}) = \sum_{r=0}^{n} B_r'^{(k)}(a \mid n+1)\sin(2r+1)z = S_{2n+1}^{2k+1}(z,a). \tag{31d}$$

According to (5a), (9), (10a), (15), (20), (25) and (30) the exact polynomial solutions of the Klein-Gordon equation (1) can be written as

$$\Phi_p = \exp[+i(\hat{p}\hat{x} + p_x x + p_z z)] \times \exp[-(a/4)\cos\xi] \times IP_n^k(\xi \mid a), \tag{32}$$

$$IP_n^k(\xi \mid a) \equiv \varphi_n^k(\xi \mid a,\mathrm{c}), \quad \varphi_n^k(\xi \mid a,\mathrm{s}), \quad \chi_n^k(\xi \mid a,\mathrm{c}) \ \text{ or } \ \chi_n^k(\xi \mid a,\mathrm{s}). \tag{32a}$$

For short, we have denoted the (Ince) polynomials (15), (20), (25) and (30), by the general symbol $IP_n^k(\xi \mid a)$. For each $n-$values (corresponding to a discrete set of transverse momenta $2p_x = (q+1)k_p$), the coefficients of the polynomials given by (15), (20), (25) and (30), satisfy the inhomogeneous equations (12d), (18c), (23c) and (28c), respectively. There are $n$ (or $n+1$) real, linearly independent eigenvectors $\boldsymbol{A}^{(k)}$, $\boldsymbol{B}^{(k)}$, $\boldsymbol{A}'^{(k)}$ and $\boldsymbol{B}'^{(k)}$. The eigenfunctions associated to the eigenvalues $\eta_n^{(k)}$ form a *doubly infinite set* labeled by the integer numbers $n$ and $k$. The eigenvalues are all real, and the eigenfunctions are real trigonometric polynomials, which coincide with the Ince polynomials, as has been explicitely displayed by (16d), (21d), (26d) and (31d). The possible values of the momentum parameter of the solutions are connected to eigenvalues $\eta_n^{(k)}$ by (10c). Since $p_x$ is given (it is either $p_x = (n + 1/2)k_p$ or $p_x = (n' + 1)k_p$, where $n = 1,2,...$ and $n' = 0,1,...$), the relation (10c) determines $n$ (or $n'$) possible values for $\hat{p}$, which may be labeled as $\hat{p}_n^{(k)}$, where 0, $1 \le k \le n$. We note that for any value of the transverse momentum $p_x = (q+1)k_p/2$ (regardless of $q$ is an integer or not) it is always possible to find a denumerably infinite set of basically-periodic even solutions $C_q^m(z,a)$ and odd solutions $S_q^m(z,a)$, associated to the eigenvalues $a_q^m(a)$ and $b_q^m(a)$, respectively (Arscott 1964). In the limit of small intensities ($a \to 0$) these transcendental Ince functions go over to $\cos mz$ and $\sin mz$.

We also note that an analogous new set of exact solutions of the corresponding Dirac equation have been recently found by us (Varró 2013), which are also expressed by finite trigonometric polynomials. However, these new polynomials are complex Fourier sums, whose interrelation with the Ince polynomials (which have an already well-developed theory) has not been explored, yet.

## 4. Numerical illustrations of some basic properties of the new exact solutions

The investigation of various possible physical applications of the exact solutions (32) is out of the scope of the present paper. The emphasize has been on proving the existence and on costructing the general polynomial solutions. In the present section we shall only briefly discuss a possible physical background of these solutions, and give few numerical illustrations.

Each of the solutions (32) contain the exponential factor $\exp[-(a/4)\cos\xi]$ which can be expanded into an infinite Fourier series (Gradshteyn and Ryzhik 1980 or Abramowitz and Stegun 1965),

$$\exp[-(a/4)\cos\xi] = \sum_{l=-\infty}^{\infty} I_l(a/4)\exp[il(\xi-\pi)] = I_0(a/4) + 2\sum_{l=1}^{\infty} I_l(a/4)\cos[l(\xi-\pi)], \qquad (33)$$

where $I_l(z)$ denote the modified Bessel functions of first kind of order $l$. This means that, though the polynomials $IP_n^k(\xi \mid a)$ are finite-term expressions, the complete wave function (32) contains all the higher-harmonics of the fundamental frequency. If $a$ is large, then this function is peaked at the points $\xi = \pi - 2k\pi$ ($k$ being an arbitrary integer) with an exponentially large contrast, as is shown in Fig.1.

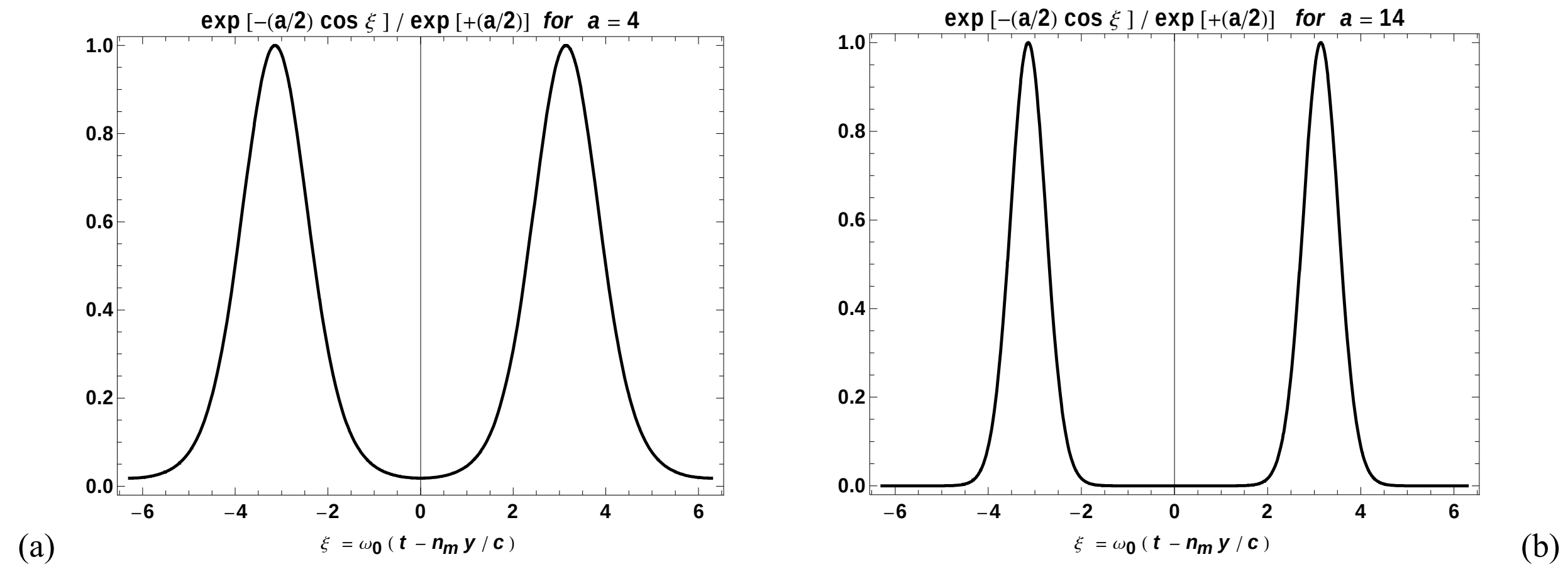

**Figure 1.** Shows the normalized square of the envelope function, $\exp[-(a/2)\cos\xi]$, in the wave function (32) for two values of the fundamental parameter $a = 4|\varepsilon|A_0/k_p$. (a): $a=4$, (b): $a=14$. At the center of the base interval $-\pi \le \xi \le +\pi$ the 'probability density' is practically zero; this void region represents a sort of 'bubble'.

Figure 1 illustrates the simple fact that the exponential factor $\exp[-(a/2)\cos\xi]$ behaves as an envelope on the base interval, rather than showing some structured modulations, which might be expected on the basis of its high hamonic content, according to (33). For large $a$ the spectrum, governed by the modified Bessel function of first kind, is practically independent of the index $l$, owing to the asymptotic behaviour $I_l(z) = (e^z/\sqrt{\pi z})[1+O(1/z)]$ (Abramowitz and Stegun 1965).

The differential equation (10a) has led to a two-parameter eigenvalue problem, where we have considered $a = 4|\varepsilon|A_0/k_p$ as a 'fundamental parameter', and $2p_x = (q+1)k_p$ and $\eta$ were 'disposable parameters'. We have seen that the condition $q+1=2n$ leads to even cosine-type and even sine-type solutions, and $q+1=2n+1$ to odd cosine-type and odd sine-type solutions. On the basis of the two equivalent forms of the eigenvalues $\eta$ in (10c), the momentum parameters can be expressed as

$$\hat{p}_n^{(k)} = \pm(k_p/2)\sqrt{\eta_n^{(k)} - (q+1)^2 - (2p_z/k_p)^2 - (2\kappa/k_p)^2 - (a/2)^2}\,, \qquad 2p_\xi/k_p^2 = \pm\sqrt{\eta_n^{(k)} - (a/2)^2}\,. \quad (34)$$

The second equation of a simpler form in (34) has been obtained by imposing (in addition) the free mass-shell condition ( $p^2 = \kappa^2$ ) for the momentum parameter $p_\mu$, where the definition of $p_\xi$ in (10c) have also been used. The eigenvalues $\eta_n^{(k)}$ are the roots of the characteristic equations $\lambda_n(\eta,a)=0$, $\mu_n(\eta,a)=0$, $\kappa_n(\eta,a)=0$ and $\tilde{\kappa}_n(\eta,a)=0$, where the characteristic polynomials have been defined in (14), (19), (24) and (29). In general, $p_0$ and $p_y$ can be expressed as

$$p_0 = k_0(p_\xi/k_p^2 + n_m\hat{p}/k_p)\,, \qquad p_y = k_0(n_m p_\xi/k_p^2 + \hat{p}/k_p)\,, \qquad (34a)$$

regardless of any constraints. If we impose the free mass-shell condition $p^2 = \kappa^2$, then, from (34), by choosing the positive sign in the second equation, we have

$$p_0 = \frac{k_0}{2}\left[\sqrt{\eta_n^{(k)} - (a/2)^2} \pm n_m \sqrt{\eta_n^{(k)} - (a/2)^2 - (q+1)^2 - (2\kappa/k_p)^2 - (2p_z/k_p)}\right], \tag{34b}$$

$$p_y = \frac{k_0}{2}\left[n_m \sqrt{\eta_n^{(k)} - (a/2)^2} \pm \sqrt{\eta_n^{(k)} - (a/2)^2 - (q+1)^2 - (2\kappa/k_p)^2 - (2p_z/k_p)}\right]. \tag{34c}$$

According to (34), the parameter $p_\xi$ ( which reduces to the energy $p_\xi / k_0 \to p_0$ in the limit $n_m \to 0$ of the pure electric field case) can only be real if $\eta \geq a^2/4$. For these eigenvalues, $p_\xi / k_0 = p_0 - n_m p_y = \pm(k_p^2/2k_0)\sqrt{\eta_n^{(k)} - a^2/4}$ *would* correspond to a sort of 'gap states', because it can well happen that $-mc^2 < \hbar c p_0 < +mc^2$ in the pure electric case. On the other hand, one has to keep in mind that, in general, the parameters $p_0$ and $p_y$ are not 'good quantum numbers' in the case of the interaction we are considering. Besides the transverse momenta $p_x = k_p(q+1)/2$ and $p_z$, only the 'longitudinal combination' $\hat{p} = (k_0/k_p)(p_y - n_m p_0)$ is a constant of motion, and $p_0$ and $p_y$ separately do not have an immediate physical meaning. Moreover, not all the eigenvalues give physically acceptable solutions of the Klein-Gordon equation. If $\hat{p}$ becomes purely imaginary, then the wave functions necessarily contain the exponentially growing factor $\exp[\pm|\hat{p}|(k_0^2/k_p^2)(y - n_m x_0)]$ in the space-like direction $\hat{k}$. This would not be an acceptable solution, *except for* the case when the interaction is limited in a finite space-time region. In the optical range $(2\kappa/k_p)^2$ is on the order of $(10^6)^2$, thus $\eta$ must be as large, in order to have a real square root and propagation constant in (34). We note that, alternatively, the laser-modified mass-shell relation $p^2 = \kappa_*^2$ may also be prescribed, where $\kappa_*^2 = \kappa^2 + \varepsilon^2 A_0^2$ contains the intensity dependent mass-shift (see e.g. Harvey *et al* 2012). In this case the first equation of (34) remains the same (because it does not depend on any restriction for the momentum parameter). The right hand side of the second equation of (34) becomes different, it is then $2p_\xi / k_p^2 = \pm\sqrt{\eta_n^{(k)}}$.

Concerning possible physical applications of the exact solutions embodied in equation (32), one has to keep in mind that, in general, the index of refraction depends on the frequencies of the spectral components of the radiation field, i.e. $n_m = n_m(\omega)$. The assumed dependence of the field amplitudes merely through the combination $\xi = k \cdot x = k^0 c(t - n_m \boldsymbol{e}_3 \cdot \boldsymbol{r}/c)$ with some constant $n_m$, is, of course, an idealization. The monochromatic field is usually ment to model a real quasi-monochromatic field having a very small spectral spread $\Delta\omega << \omega_0$ around some central frequency $\omega_0 = ck_0$, and in the present paper we assume that the associated variation of the index of refraction is negligible in this spectral range, i.e. $\Delta n_m << n_m$. For example, our analysis may be relevant for the description of the interaction of a relativistic electron with a strong laser field propagating in an underdense plasma. In the case, when the Drude free electron model applies, the index of refraction may well be approximated by the formula $n_m^2(\omega) = 1 - \omega_p^2/\omega^2$, with $\omega_p^2 = 4\pi n_e e^2/m$, where $\omega_p$ is a real plasma frequency, where $n_e$ is the density of the plasma-electrons. If $\omega > 2^{1/2}\omega_p$, then $\Delta n_m / n_m < \Delta\omega/\omega$ is satisfied for all such frequencies, thus in case of $\Delta\omega/\omega << 1$ the relative spectral variation of the index of refraction is also negligible, $\Delta n_m / n_m << 1$. Under these conditions, the vector potential displayed in (2) can be a faithful representation of a quasi-monochromatic radiation field in a homogeneous plasma medium. The invariant square (effective mass of the EM field) of the propagation vector $k_\mu k^\mu = k_0^2(1 - n_m^2(\omega)) = \omega_p^2/c^2$ does not depend on the circular frequency $\omega$ of the laser field.

The fundamental parameter $a$ can be expressed in terms of various combinations of parameters characterizing the applied monochromatic field and the medium. Taking the free electron model of a plasma medium, with a dielectric permittivity $\varepsilon_m(\omega) = 1 - \omega_p^2/\omega^2 = n_m^2(\omega)$, the electromagnetic plane wave under discussion has the dispersion relation $\omega(k_y) = \sqrt{\omega_p^2 + (ck_y)^2}$, with $k_y = n_m\omega/c$, corresponding to the phase $\xi = \omega(t - n_m y/c)$ of the laser field. In this case the parameter $a$ can be written as the work done by the electric force along the plasma wavelength divided by the photon energy. The ratio of the photon density and the electron density also naturally appears,

$$a = 4\varepsilon A_0 / k_p = 4\frac{eF_0 \lambda\!\!\!^{-}{}_p}{\hbar\omega_0} = 4\sqrt{(2mc^2/\hbar\omega_0)(n_{ph}/n_e)} = 2\mu_0(2mc^2/\hbar\omega_p) \ , \tag{35a}$$

$$n_{ph} \equiv \frac{I_0}{c\hbar\omega_0} = 2.08\times10^8 \times (S/E_{ph})[cm^{-3}], \qquad \mu_0 \equiv \frac{eF_0}{mc\omega_0} = 1.06\times10^{-9}\times S^{1/2}/E_{ph} \ , \tag{35b}$$

$$\omega_p = \sqrt{4\pi n_e e^2/m} \ , \quad k_p = \omega_p / c = 1/\lambda\!\!\!^{-}{}_p = 2\pi/\lambda_p \, . \tag{35c}$$

We have used the definition $k_p = k_0\sqrt{1 - n_m^2}$ of the 'plasma wavenumber' in (10b), and introduced the true plasma frequency $\omega_p$. For example, the electron density $n_e = 7.242\times10^{20}cm^{-3}$ corresponds to the plasmon energy $\hbar\omega_p = 1eV$ (in which case the plasma wavelength equals to $\lambda_p = 1240nm$). In the numerical expressions of the photon number density $n_{ph}$ and of the well-known 'dimensionless intensity parameter' $\mu_0$ (which nowadays also called 'scaled vector potential'), the quantities $S = I_0/(W/cm^2)$ and $E_{ph} = \hbar\omega_0 /(eV)$ measure the intensity in Watts per square centimeters and the photon energy in electron Volts, respectively. For a Ti:Sa laser with $\lambda_0 = 795nm$ we have $\hbar\omega_0 = 1.563eV > \hbar\omega_p$, which means that this radiation freely propagates in the underdense plasma, with the phase velocity $c/n_m$ and group velocity $cn_m$. In this case $a = 2\times10^6 \mu_0$ and $\mu_0 = 6.782\times10^{-10} \times S^{1/2}$, showing that for optical frequencies the parameter $a$ is by 6 orders of magnitude larger than the usual intensity parameter $\mu_0$.

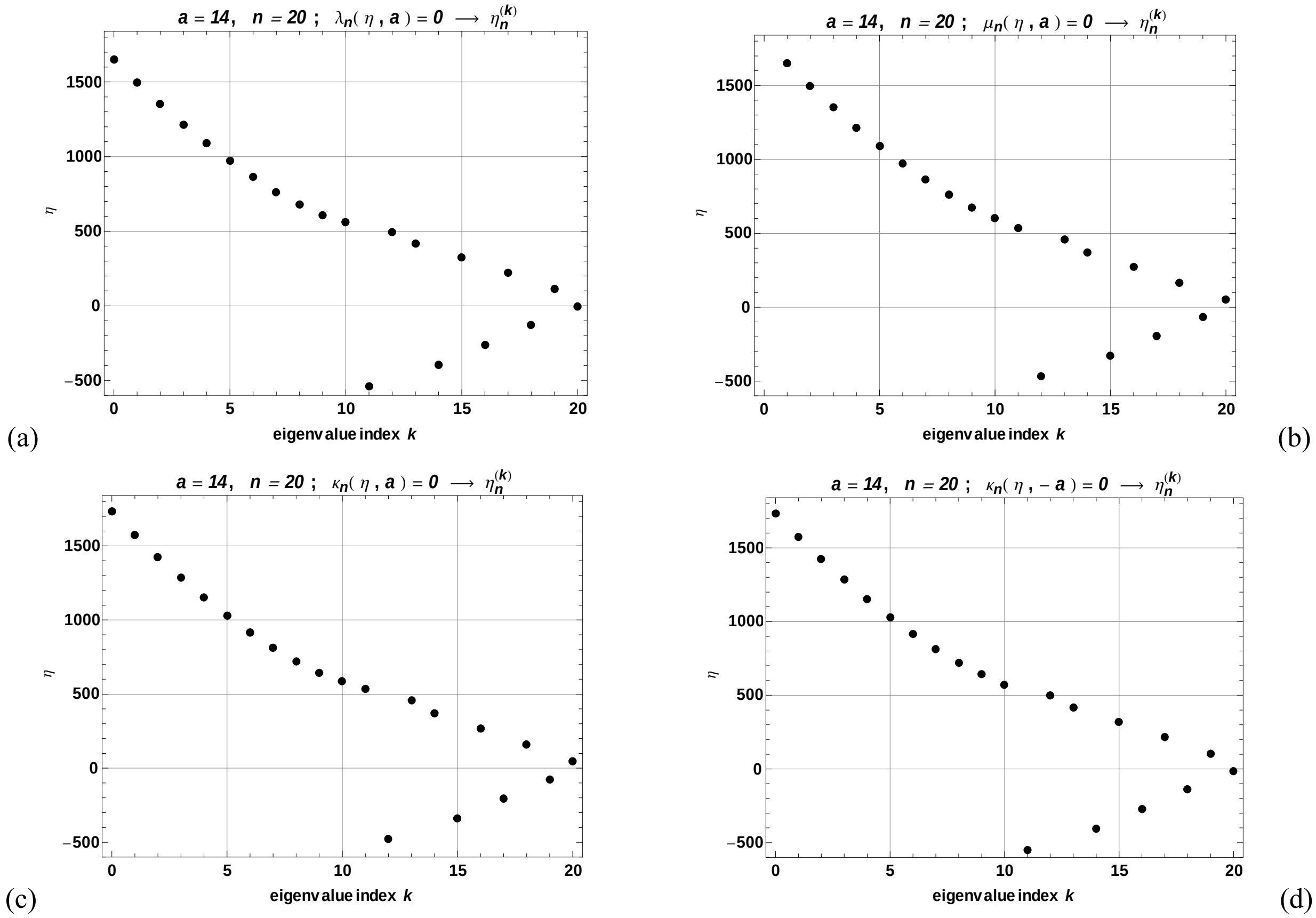

**Figure 2.** Shows the eigenvalues associated to the four types of solutions (32a), for $a = 14$ and $n = 20$. We have taken $\hbar\omega_p = 1eV$ and $n = 20$, i.e. the transverse momentum of the electron (in original units) is $p_x = 20\times\hbar k_p$, where $k_p = \omega_p / c$. The interaction with a Ti:Sa laser field of photon energy $\hbar\omega_0 = 1.563eV$ and peak intensity $I_0 = 100MW/cm^2$ has been considered as an example. The eigenvalues in (a), (b), (c) and (d) are the roots of the characteristic equations $\lambda_n(\eta,a) = 0$, $\mu_n(\eta,a) = 0$, $\kappa_n(\eta,a) = 0$ and $\tilde{\kappa}_n(\eta,a) = 0 = \kappa_n(\eta,-a)$, where the characteristic polynomials have been defined in (14), (19), (24) and (29), respectively. Notice that in (a) there are 20 eigenvalues, and in (b), (c) and (d) they are 21. We have not used Ince's ordering of the eigenvalues displayed in (16a), (21a), (26a) and (31a). For instance, the smallest (negative) eigenvalues in (a) have the indexces $k = 11$, $k = 15$, $k = 17$ and $k = 19$ in our figure, they would be labeled by $m = 0$, $m = 1$, $m = 3$ and $m = 4$, according to Ince's convention. The largest eigenvalue has index $k = 0$, which corresponds to the largest $m = 21$.

Even for a relatively moderate intensity of $100MW/cm^2$ $(S=10^8)$, the value of $a=13.56>>1$ is already quite large. In Figure 1 we present illustrations for the eigenvalue spectrum in case of such a moderate intensity, for which we assumed $a=14$, and have taken $n=20$. In a recent experiment Kiefer *et al* (2013) have used a Ti:Sa laser of intensity $I_0=6\times10^{20}W/cm^2$, in which case $\mu_0=16.61$ is well beyond the relativistic limit ($\mu_0=1$), and consequently $a=3.32\times10^7$ is enormously large. We note that in such extreme cases there exist approximate analytic formulae for both the eigenvalues and for the Ince polynomials, however, in the present paper we do not have space to enter into these details.

In Figure 3 we show the distribution of the harmonic strengths (like $[A_r^{(k)}(a\,|\,n+1)]^2$ from (15)) in the expansion of the polynomial factors, in case of the interaction with a Ti:Sa laser field of photon energy $\hbar\omega_0=1.563eV$ and peak intensity $I_0=100MW/cm^2$ in a plasma medium of electron density $n_e=7.242\times10^{20}cm^{-3}$ ($\hbar\omega_p=1eV$). Of course, the complete physical spectrum may in principle be influenced by the exponential prefactor leading to the modified Bessel function series in (33), but in the present paper we would like to put the emphasize on the properties of the less-known polynomial factors. According to Figure 3, these spectra are in general qualitatively different from each other. As the index $k$ of the eigenvalue increases, the single-peaked structure gradually shifts from the higher harmonics to the lower harmonics. When the first negative eigenvalue is reached, then the harmonic distribution becomes oscillatory, and beyond this border this oscillations are also present in the solutions belonging to positive eigenvalues. In Figure 4 we give an overview of the harmonic spectra in Figure 3.

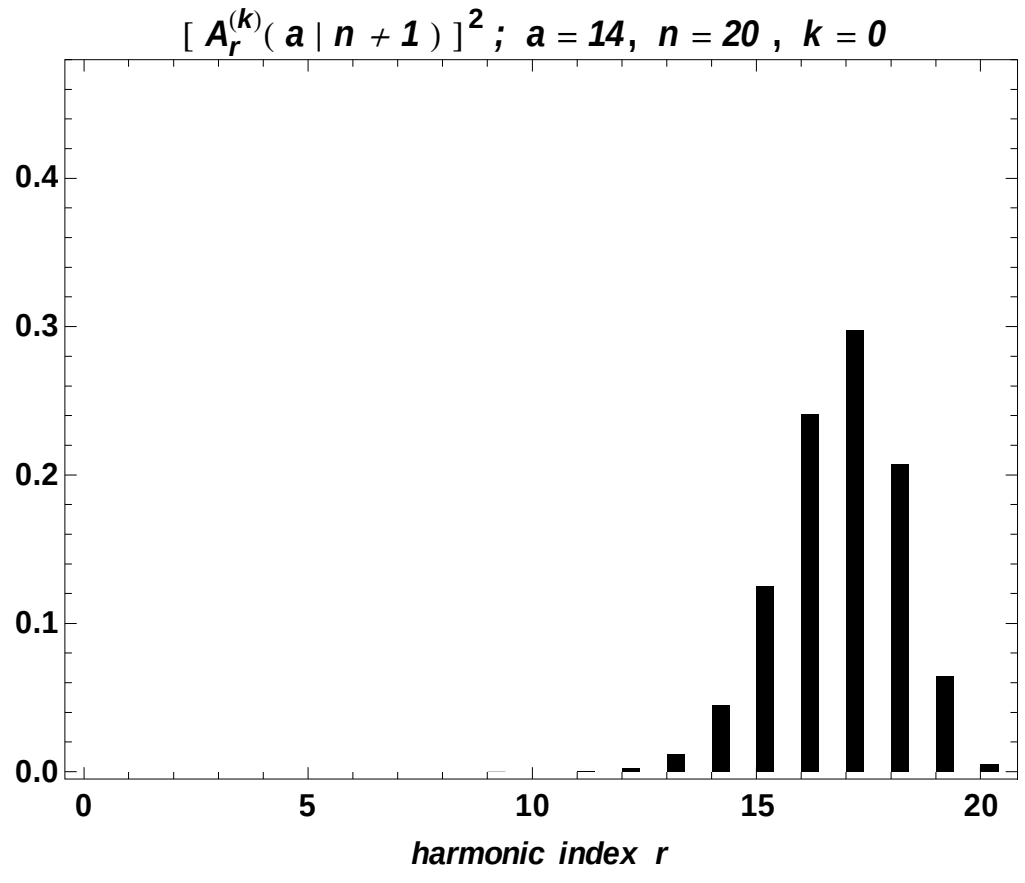

(a)

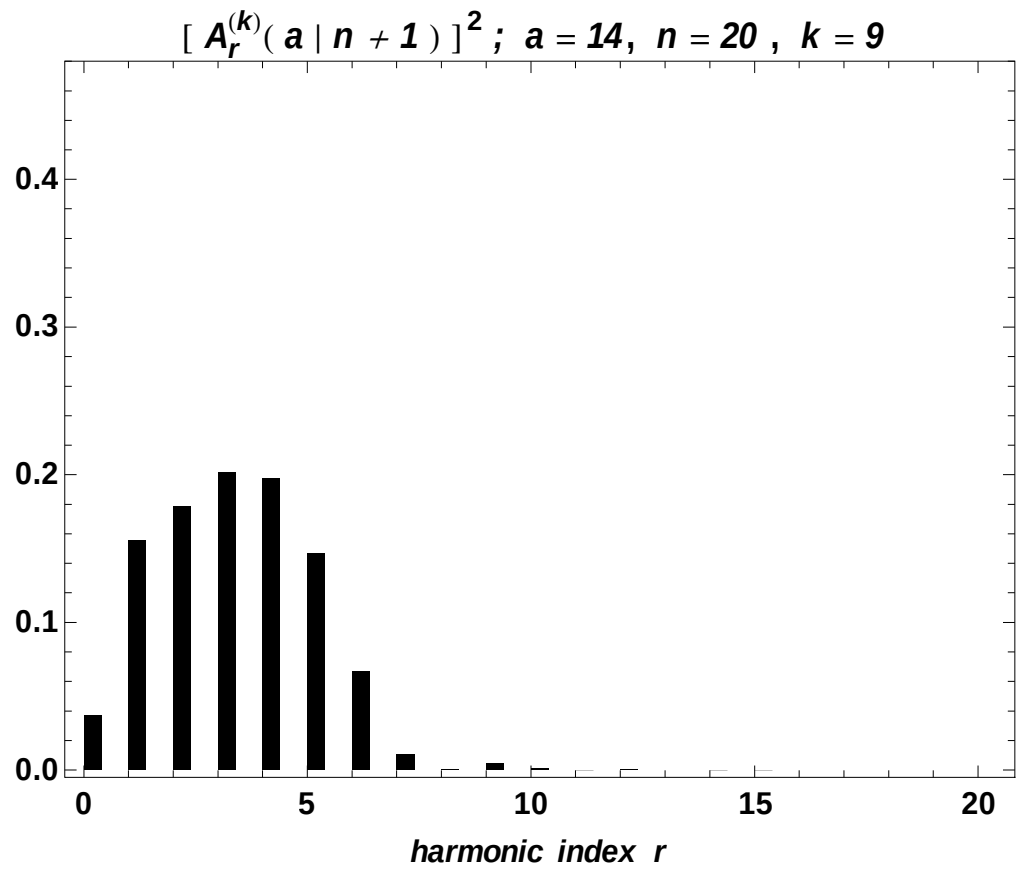

(b)

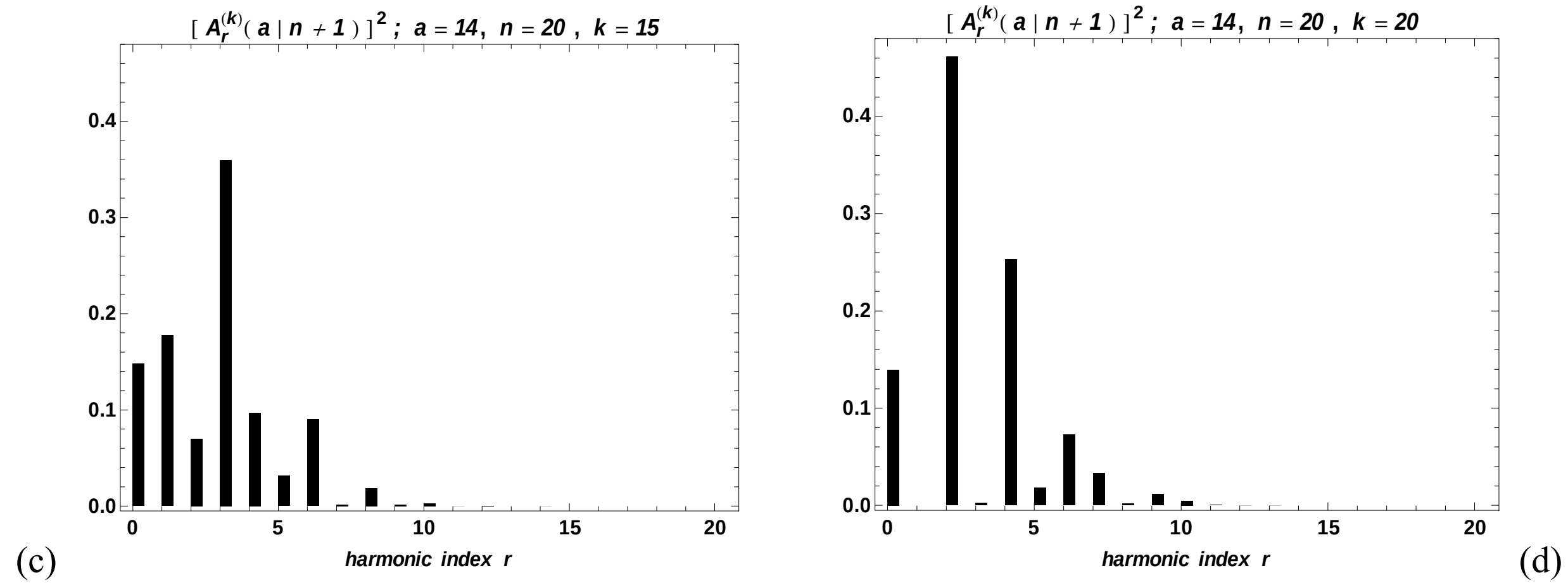

**Figure 3.** Shows the strength $[A_r^{(k)}(a\,|\,n+1)]^2$ of the harmonic coefficients of the polynomial $\varphi_n^k(\xi\,|\,a,\mathrm{c})$ for four eigenvalues labelled by the upper index $k = 0,\ 9,\ 15,\ 20$. We are considering the case when $a = 14$ and $n = 20$, and the input parameters are the same as in Figure 2.

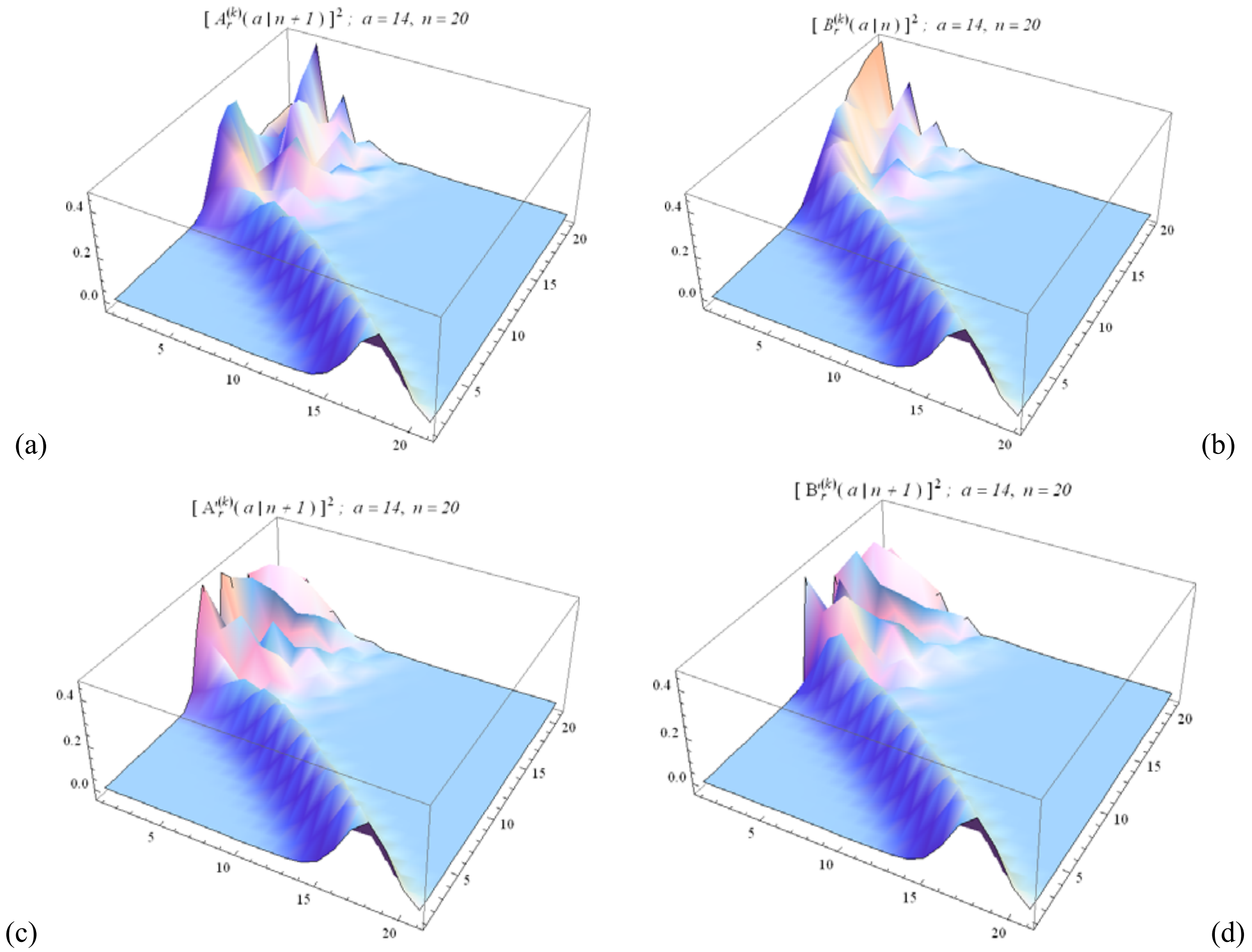

**Figure 4.** Shows an overview of the strength of the harmonic coefficients on a three-dimensional list plot when $a = 14$ and $n = 20$. (a): $[A_r^{(k)}(a\,|\,n+1)]^2$, (b): $[B_r^{(k)}(a\,|\,n)]^2$, (c): $[A_r'^{(k)}(a\,|\,n+1)]^2$ and (d): $[B_r'^{(k)}(a\,|\,n+1)]^2$. The input parameters are the same as in Figure 2. This figure summarizes the behaviour of these quantities as functions of the harmonic numbers ($r = 0, 1, 2, ..., 20$ on the left axis), for different eigenvalues ($k = 0, 1, 2, ..., 20$, which are drawn on the right axis). The discrete points are connected by a smoothened surface in order to guide the eye. For the lowest index $k = 0$ (or $k = 1$ in (b)) the distribution is concentated to large $r-$values (left axis) in a single peak. For medium $k-$values a double peak structure develops, and for larger values goes over to an oscillatory distribution.

In Figure 5 we illustrate the detailed temporal evolution of the even cosine-type wave functions, whose harmonic spectra have been shown in Figure 3. Finally, Figure 6 illustrates the temporal evolution of the even sine-type wave functions The further numerical study of the exact solutions (32) is out of the scope of the present paper.

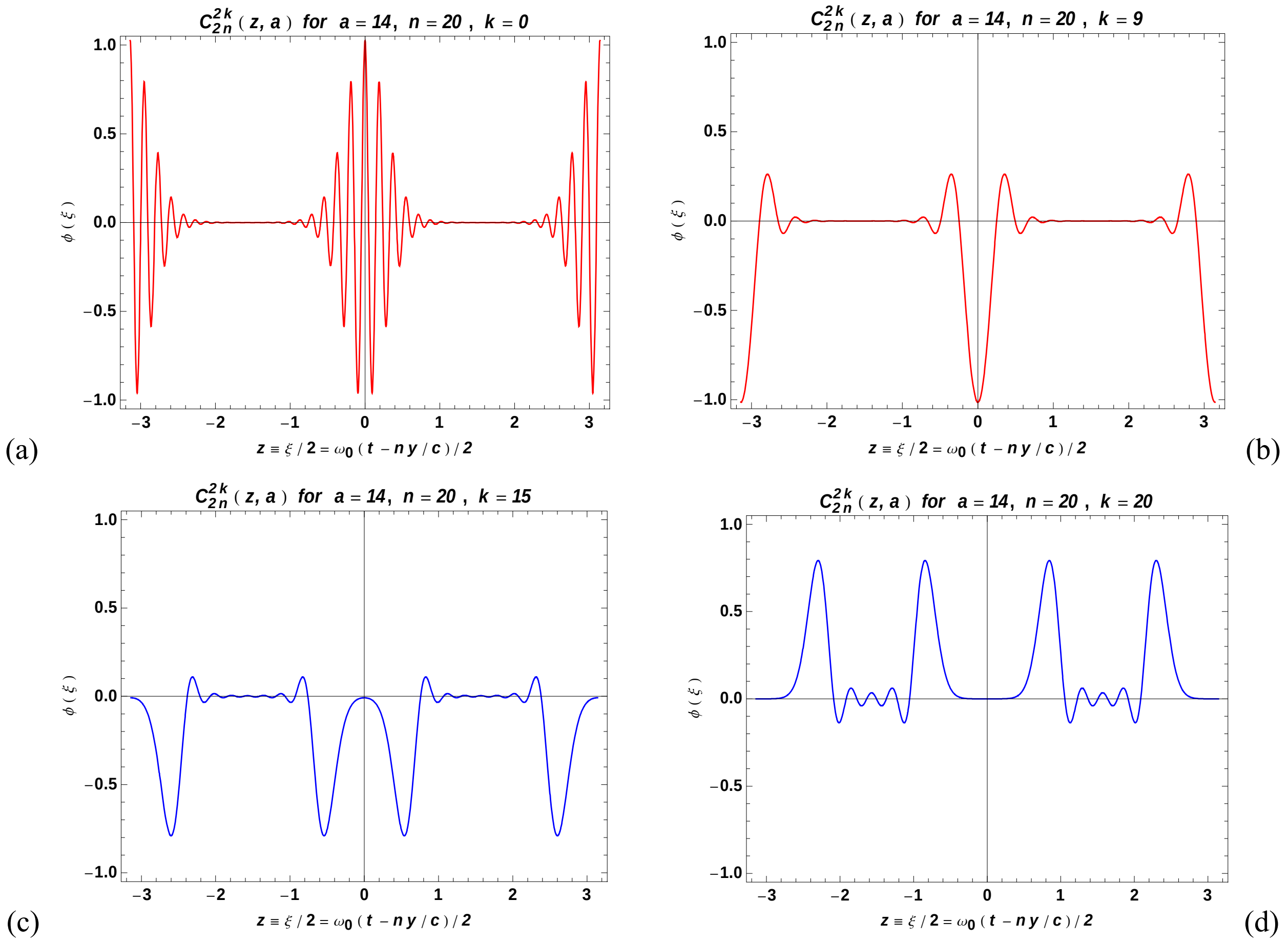

**Figure 5.** Shows the shape of the even cosine-type polynomial solutions $\varphi_n^k(\xi \mid a, \mathrm{c})$ for four eigenvalues labelled by the upper index $k = 0, 9, 15, 20$. $a = 14$ and $n = 20$, and the input parameters are the same as in Figure 2. These functions are finite sums of the cosine function $\cos 2rz = \cos r\xi$, and their harmonic spectra of these functions have been shown in Figure 3.

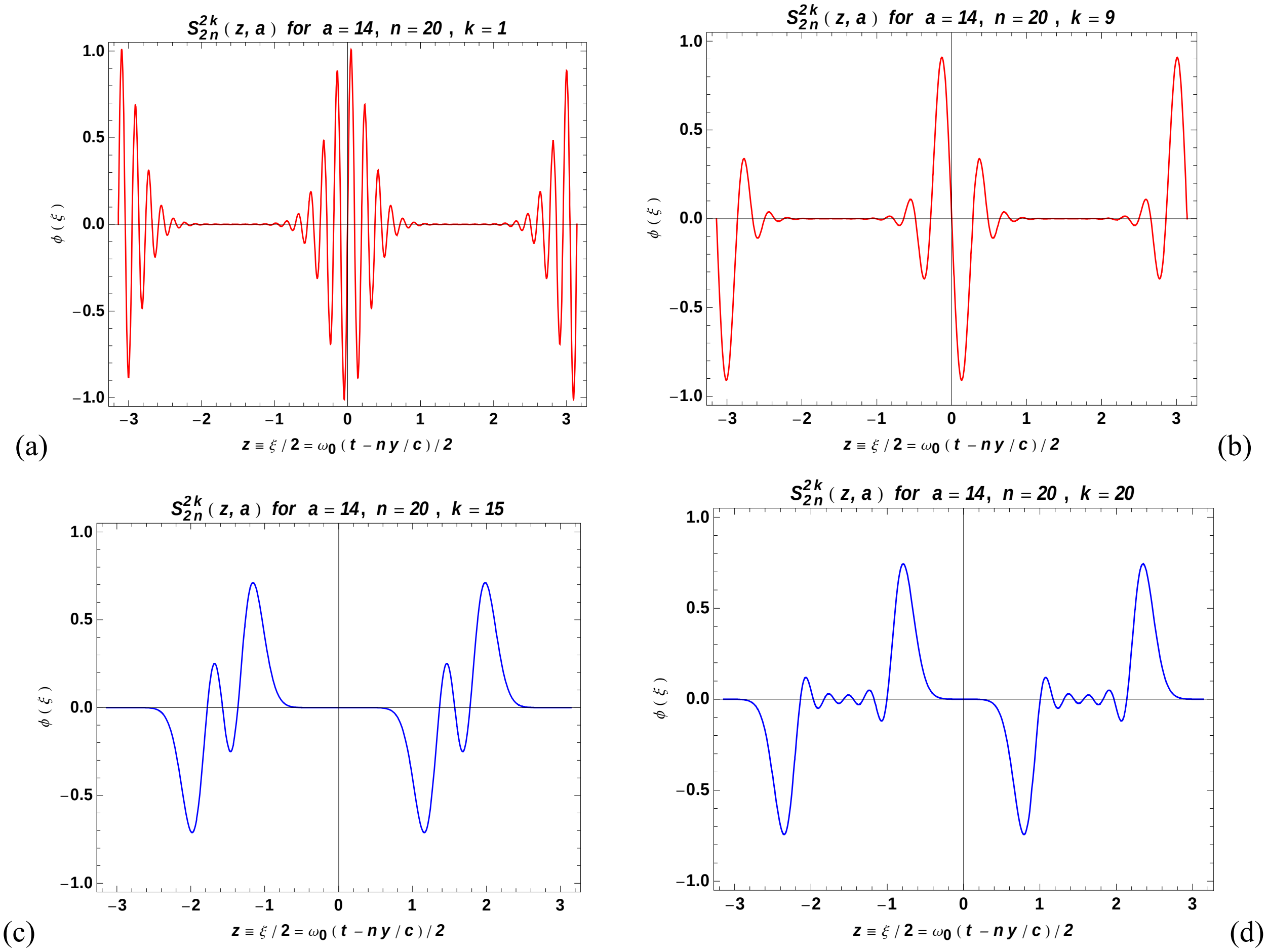

**Figure 6.** Shows the shape of the even sine-type polynomial solutions $\varphi_n^k(\xi \mid a, s)$ for four eigenvalues labelled by the upper index $k = 1,\ 9,\ 15,\ 20$ . $a = 14$ and $n = 20$ , and the input parameters are the same as in Figure 2. These functions are finite sums of the sine function $\sin 2rz = \sin r\xi$ , as shown in (20), (21d) and in the summarizing equations (32), (32a)

## 5. Summary

We have presented closed form exact solutions of the Klein-Gordon equation of a charged particle (electron) moving in a monochromatic classical plane electromagnetic field in a medium of index of refraction $n_m < 1$. The solutions have been expressed by the Ince polynomials which form a doubly infinite discrete set, labeled by two integer quantum numbers. Physically, they represent quantized spectra of the electron's momentum components along the polarization and along the propagation direction of the laser field. The odd Ince polynomials are finite sums of cosines or sines of the half-integer multiples of the phase of the incoming field, thus they describe half-integer order higher harmonics. We note that in mathematical physics Ince polynomials also appear in the description of various other systems. For instance, if one separates the two-dimensional harmonic oscillator wave function by using elliptic coordinates (Boyer et al 1975) the eigenfunctions are products of Ince polynomials.

As an example for a physical interpretation of our solutions, we have considered the medium as an underdense plasma. In this case one of the quantum numbers $n$ characterizes the transverse momentum $p_x = nk_p$ or $p_x = (n + \frac{1}{2})k_p$ of the test electron, where $k_p = \omega_p / c$, with $\omega_p$ being the plasma frequency. The other quantum number determines the possible values of the longitudinal energy-momentum parameter. The fundamental parameter $a$ (see (35a)) which determines the strength of the interaction is the work done on the electron by the electric force of the laser field along a plasma wavelength divided by the photon energy. This $a$ is a quantum parameter, and it is typically by many orders of magnitudes larger than the intensity parameter $\mu_0$ (see (35b)), which usually appears in the theory of strong-laser field matter interactions. The existence of the periodic 'bound states' we have found, may have relevance concerning possible quantum features of mechanisms of laser acceleration of electrons by high-intensity laser fields and wake-fields in an underdense plasma (see e.g. Xia *et al* 2012, Mourou *et al* 2013).

In secion 2 we have derived the relevant Ince equation from the original Klein-Gordon equation, and made a brief comparison with the derivation of the usual Volkov states, and of the Mathieu-type wave functions in a medium. In section 3 we have derived the two even and two odd periodic, finite solutions, which coincide with the corresponding Ince polynomials. These solutions are are associated to the eigenvalue problem of finite special tri-diagonal matrices (13a-b) and (16a-b). We have restricted our analysis only to these special class of solutions, proportional with polynomial expressions. They form a subset of all solutions of the Ince equation, including the transcendental Ince functions. In section 4 we have given some numerical examples for the relevant parameters appearing in the analysis, and presented numerical illustrations for the discrete eigenvalues, the temporal shape of the wave functions, and for the harmonic spectra which are associated to them.

## Acknowledgments

This work has been supported by the Hungarian Scientific Research Foundation OTKA, Grant No. K 104260.